\documentclass[journal, 10pt, final]{IEEEtran}

\usepackage{epsfig}
\usepackage{graphicx}
\usepackage{amsmath}
\usepackage{amssymb}

\usepackage{xcolor}
\usepackage[english]{babel}
\usepackage[utf8]{inputenc}
\usepackage{mathrsfs}
\usepackage{booktabs}
\usepackage{multirow}

\ifCLASSOPTIONcompsoc
\usepackage[caption=false,font=normalsize,labelfon
t=sf,textfont=sf]{subfig}
\else
\usepackage[caption=false,font=footnotesize]{subfig}
\fi

\usepackage{ifthen}

\newcommand{\myarrowhead}{{Latex[length=1mm,width=1mm]}} % use instead of "<" or ">" as arrowhead

\newcommand{\TT}{\ensuremath\mathrm{T}}
\newcommand{\ee}{\ensuremath\mathrm{e}}
\newcommand{\ii}{\ensuremath\mathrm{i}}
\newcommand{\CONV}{\ensuremath\mathrm{\;**\;}}
\newcommand{\MOD}[2]{\ensuremath{{#1}\,\mathrm{mod}\,{#2}}}
\newcommand{\COLVEC}[3]{\ensuremath{\begin{pmatrix}{#1}\\{#2}\\{#3}\end{pmatrix}}}

\newcommand{\tikzsize}{\footnotesize}

\newcommand{\myvec}[1]{\boldsymbol{\mathbf{#1}}}
\newcommand{\norm}[1]{\ensuremath{\lVert #1 \rVert}}

\newcommand{\CHO}{Cho et al.}
\newcommand{\DANS}{Dansereau et al.}	
\newcommand{\MLTB}{\textsc{Matlab} \textit{Light Field Toolbox}}

\newcommand{\FIG}[1]{Fig.~\ref{#1}}
\newcommand{\TAB}[1]{Tab.~\ref{#1}}

\definecolor{myblue30}{rgb}{0.78, 0.82, 0.91}
\definecolor{myblue15}{rgb}{0.91, 0.93, 0.96}
\definecolor{seaborn-blue}{rgb}{0.2823529411764706, 0.47058823529411764, 0.8156862745098039}
\definecolor{seaborn-blue15}{rgb}{0.8502, 0.8765, 0.8911}
\definecolor{seaborn-blue10}{rgb}{0.9002, 0.9177, 0.9274}

\definecolor{seaborn-red}{rgb}{0.8352941176470589, 0.3686274509803922, 0.0}
\newcommand{\COLORONE}{seaborn-blue}
\newcommand{\COLORTWO}{seaborn-red}

\newcommand{\olit}[2][3]{{}\mkern#1mu\overline{\mkern-#1mu#2}}

\newcommand{\laplacez}{\mbox{\setlength{\unitlength}{0.1em}% zweidimensionales Hantelsymbol Hintransformation
		\begin{picture}(20,10)%
		\put(2,3){\circle{4}}%
		\put(4,2.3){\line(1,0){13}}%
		\put(4,3.7){\line(1,0){13}}%
		\put(18,3){\circle*{4}}%
		\end{picture}%
	}%
}%

\newcommand{\TZz}{\hspace{2mm}\laplacez\hspace{2mm}} % Hantelsymbol Hintransformation
\newcommand{\vTZz}{\hspace{.75ex}\rotatebox[origin=c]{270}{\laplacez}\hspace{.75ex}}

\usepackage[pagebackref=true,breaklinks=true,letterpaper=true,colorlinks,bookmarks=false]{hyperref}

\usepackage{siunitx}
\usepackage{textcomp}
\begin{document}

%%%%%%%%% TITLE
%\title{Pre-calibration of hand-held plenoptic cameras: A quantitative approach to microlens array grid estimation and its influence on light field decoding and calibration}
\title{Microlens array grid estimation, light field decoding, and calibration}
%\title{A quantitative approach to evaluate the pre-calibration pipeline of micro lens array based cameras.}

\author{
	Maximilian Schambach\thanks{© 2020 IEEE.  Personal use of this material is permitted.  Permission from IEEE must be obtained for all other uses, in any current or future media, including reprinting/republishing this material for advertising or promotional purposes, creating new collective works, for resale or redistribution to servers or lists, or reuse of any copyrighted component of this work in other works.} and Fernando Puente Le\'{o}n, \textit{Senior Member, IEEE}\\
	Karlsruhe Institute of Technology, Institute of Industrial Information Technology\\
	Hertzstr. 16, 76187 Karlsruhe, Germany\\
	{\tt\small \{schambach, puente\}@kit.edu}
	\vspace{-0.7cm}
}

\maketitle
%\thispagestyle{empty}

%%%%%%%%% ABSTRACT
\begin{abstract}
   We quantitatively investigate multiple algorithms for microlens array grid estimation for microlens array-based light field cameras. Explicitly taking into account natural and mechanical vignetting effects, we propose a new method for microlens array grid estimation that outperforms the ones previously discussed in the literature.
   To quantify the performance of the algorithms, we propose an evaluation pipeline utilizing application-specific ray-traced white images with known microlens positions.  Using a large dataset of synthesized white images, we thoroughly compare the performance of the different estimation algorithms.
   As an example, we apply our results to the decoding and calibration of light fields taken with a Lytro Illum camera. We observe that decoding as well as calibration benefit from a more accurate, vignetting-aware grid estimation, especially in peripheral subapertures of the light field.
\end{abstract}

%%%%%%%%% BODY TEXT

\begin{figure*}[ht]
	\centering
	\hspace{1mm}
	\subfloat[]{\includegraphics{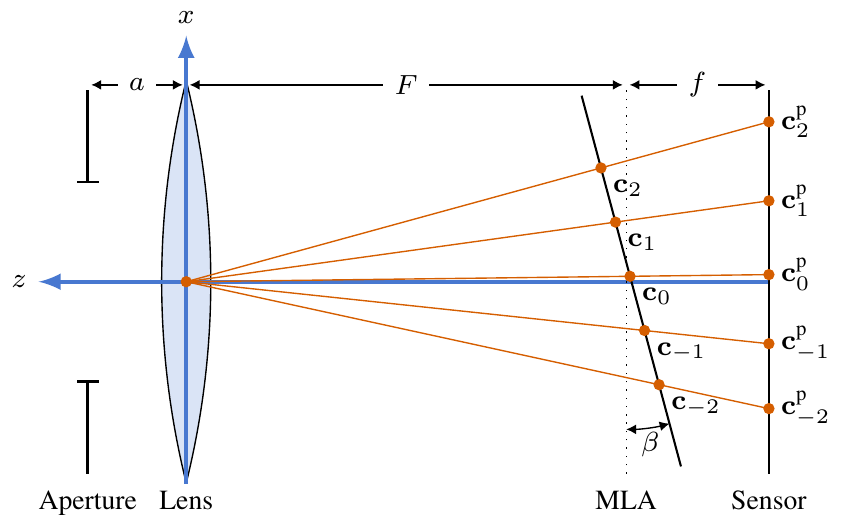}} 
	\hfill
	\subfloat[]{\includegraphics{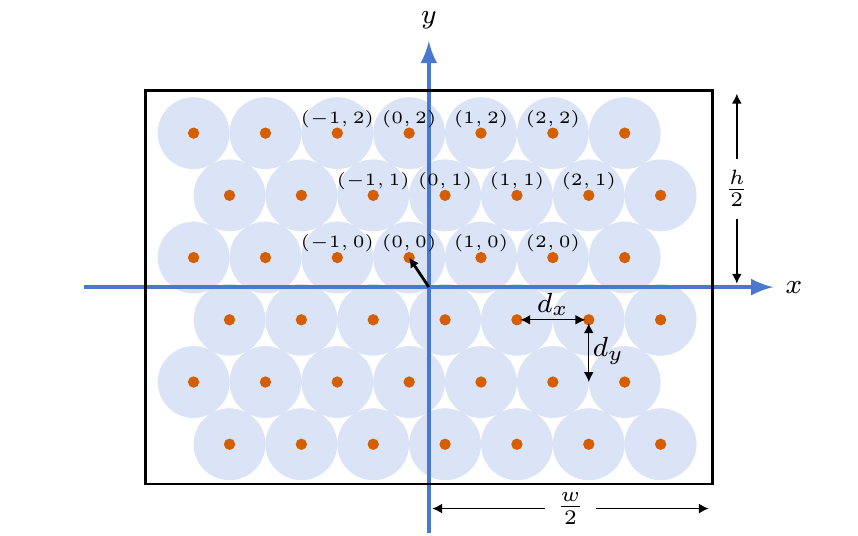}}
	\caption{Schematic drawings of the used camera model of the unfocused plenoptic camera with exaggerated ML size. (a) A 2D section ($y = 0$) of the used camera model. The coordinates $\myvec{c}_{\pm i}$ denote the centers of the ML (which are not explicitly depicted) and their perspective projections $\myvec{c}_{\pm i}^{\textrm{p}}$. (b) The used MLA model before rotation and tilt. The ML centers $\myvec{c}_{ij}$ are depicted in orange with corresponding index labels $(i, j)$.}
	\label{fig:camera-model}
\end{figure*}

%=======================================================================
\section{Introduction}
%=======================================================================
Computational cameras, that is, cameras utilizing combined optical and digital image processing techniques, have been gaining attention both in consumer applications, such as multi-lens camera systems in mobile devices, as well as scientific and industrial applications, such as light field cameras \cite{Adelson1992, Ng2005} or snapshot hyperspectral cameras \cite{Arce2014}. Computational imaging systems can usually well be described using the so-called \emph{4D light field} $L_{\lambda, t}(u, v, a, b)$, where $\lambda$ describes a wavelength, $t$~the time, and the coordinates $(u, v, a, b)$ correspond to a certain parametrization of the spatio-angular dependency of the light field of which there are numerous. For computational cameras, one usually uses the plane-plane parametrization: a light ray inside a camera is uniquely described by the intersection points $\myvec{u} = (u, v)$ and $\myvec{a} = (a, b)$ of two parallel planes, e.g.\ the main lens plane and the sensor plane. 

In particular, microlens arrays (MLAs) are used in computational imaging applications allowing for a complex coding (or multiplexing) scheme of the light field onto an imaging sensor. Most prominently, MLAs are used in compact MLA-based light field cameras \cite{Ng2005, Lumsdaine2009}, but also in other applications such as multi- or hyperspectral imaging \cite{Shogenji2004, Stanley2017}. As usual, a certain camera model is then used to calibrate the intrinsic (and extrinsic) parameters of the camera to relate the image-side to the object-side light field. The model is then evaluated, e.g.\ by using the ray re-projection error. In most light field camera models \cite{Dansereau2013, Cho2013, Zhang2016, Zhang2018}, the calibration is performed using image-side light fields that have been decoded from the sensor lenslet images, while others perform calibration utilizing the raw images directly \cite{Bok2017}. In either case, this includes multiple non-trivial pre-processing steps, such as the detection of the projected microlens (ML) centers and estimation of a regular grid approximating the centers, alignment of the lenslet image with the sensor, slicing the image into a light field and, in the case of hexagonal MLAs, resampling the light field onto a rectangular grid. These steps have a non-negligible impact on the quality of the decoded light field and camera calibration. Hence, a quantitative evaluation is necessary where possible. Here, we will focus on the estimation of the MLA grid parameters (to which we refer to as \emph{pre-calibration}), which is the basis for all decoding and calibration schemes found in the literature. In spite of the importance of the pre-calibration pipeline, the literature focuses mostly on the camera models and decoding but pays little or no attention to the necessary details emerging in the pre-calibration, most importantly non trivial effects such as mechanical and natural vignetting. While for a correct pre-calibration, the detection of the perspectively projected ML centers is necessary, all methods proposed in the literature rely on estimating the center of each ML image brightness distribution, approximating the orthogonally projected centers. Due to natural and mechanical vignetting, this results in severe deviations from the true projected centers, in particular in off-center MLs. This is the main scope of this article. In particular, our contributions are as follows:
\begin{itemize}
	\item We propose a camera model and ray tracer implementation to synthesize application-specific white images with known ML centers as reference data.
		
	\item We propose a new pre-calibration algorithm motivated by our physical camera model, taking into account natural and mechanical vignetting effects.
	
	\item We present detailed accuracy requirements that the pre-calibration pipeline has to fulfill and show that the proposed algorithm, in case the of a Lytro Illum camera, fulfills these requirements. We compare our algorithm to different schemes proposed in the literature (which we show fail the accuracy requirements). 
	
	\item We evaluate the full light field decoding pipeline for the different pre-calibration algorithms for simulated as well as real light field data.
	
	\item We investigate the influence of the quality of the pre-calibration on the full calibration of a Lytro Illum light field camera for different calibration methods.
\end{itemize}
Since the ML grid parameters are application-specific, we make the source code for a full evaluation pipeline (image synthesis, ML grid estimation, and decoding) freely available \cite{git} to be used in scientific research for any kind of MLA-based application. This includes the release of an open source Python framework for light field decoding and analysis as well as for applications in hyperspectral imaging. Even though the following presentation relies on a light field camera in the unfocused design \cite{Ng2005}, the proposed pre-calibration method is equally applicable to cameras in the focused design \cite{Lumsdaine2009}. 

The remainder of this paper is organized as follows. We introduce the used camera model and chosen camera parameters in Section~\ref{sec:reference-data}. In Section~\ref{sec:ml-detection}, we review different methods for the estimation of ML grids and formulate precise accuracy requirements which the algorithms ought to fulfill. Furthermore, we introduce a new estimation method which we thoroughly motivate using the physical camera model. All estimation methods are then quantitatively evaluated. In the remaining Sections~\ref{sec:lf-decoding} and \ref{sec:lf-calibration}, we investigate the influence of the ML grid estimation accuracy on light field decoding and calibration, respectively.

%=======================================================================
\section{Camera Model and Reference Data}\label{sec:reference-data}
%=======================================================================

The pre-calibration of MLA-based cameras is usually performed using so-called white images (WIs)---images of a white scene, for example taken using an optical diffuser. In order to quantitatively evaluate the performance of the ML grid estimation algorithms, appropriate reference data is needed. Of course, real WIs, as for example provided by the Lytro cameras, are unsuited since the actual ML centers are unknown. Therefore, reference data has to be synthesized. Previously, Hog et al.\ \cite{Hog2017} used a simple addition of three 2D cosine waves to synthesize a WI with known parameters, but the results are too crude for a precise evaluation of the estimation algorithms. In particular, they neither account for natural, nor mechanical vignetting of the main lens and the MLs. Here, we use a self-developed ray tracer \cite{Nuernberg2019, git}, which we have extended by the following camera model to render a multitude of white images with precisely known ML centers. 

\subsection{Camera model}
The camera model used is depicted in \FIG{fig:camera-model}. In our model, the camera consists of a main lens and a collection of MLs, arranged in a hexagonal grid, which may be rotated (not depicted in the figure) and tilted. All lenses are modeled as thin lenses. As is usual in the focused design, f-number matching of main lens and MLs is assumed. To simulate irregularities of the grid, we add independent uncorrelated Gaussian noise $\myvec{\epsilon}$ to the ideal grid point's $x$- and $y$-coordinates. Natural vignetting is implemented by using the $\cos^4 \Phi$ law and the ray's incident angle $\Phi$. Finally, an object-side aperture with variable entrance pupil is placed at distance $a$ to the main lens to account for mechanical vignetting effects. Note, that we do not model systematic, non-rigid deformations of the MLA as considered in \cite{Pitts2014}. We argue that these irregularities should be eradicated in the manufacturing process of high-quality MLAs as they introduce irreducible blur in the light field (on which we will elaborate in Section~\ref{sec:accuracy-estimates}).

The ideal, unrotated, untilted and unshifted ML center coordinates are given by
\begin{equation}
\myvec{c}_{\pm i \pm j}^{\textrm{id}} = \myvec{o}_\textrm{g} +
\begin{pmatrix}
\left( \pm i \pm \frac{1}{2} \MOD{j}{2} \right) d_x \\
\pm j d_y\\
0
\end{pmatrix} + \myvec{\epsilon}_{\pm i\pm j}\,,
\end{equation}
for $(i, j) \in \mathbb{N}^2$. Here, $d_x, d_y$ denote the ideal grid spacing, $\myvec{o}_\textrm{g} = (o_{\textrm{g}, x}, o_{\textrm{g}, y}, 0)^\TT$ the grid offset, and $\myvec{\epsilon} = (\epsilon, \epsilon, 0)^\TT$ the grid noise with variance $\sigma^2_\textrm{g}$ . The ideal hexagonal grid is determined by a single grid spacing $d$ via
$d_x = d$, $	d_y = \sqrt{3}\,d / 2$, where the ML radius is  given by $r = d / 2$. The ideal grid points are then rotated in the $xy$-plane by $\alpha$, rotated around the $y$-axis by $\beta$, rotated around the $x$-axis by $\gamma$, and shifted to $z = -F$, where $F$ denotes the main lens focal length (or image distance). Hence, we obtain the final grid point coordinates
\begin{equation}\label{eq:ml-centers}
\myvec{c}_{\pm i \pm j} = \myvec{R}_{x, \gamma}\,\myvec{R}_{y, \beta}\,\myvec{R}_{z, \alpha}\myvec{c}_{\pm i \pm j}^{\textrm{id}} + (0, 0, - F)^\TT \,.
\end{equation}
We will at times refer to them simply by $\myvec{c}_{k}$ for $k \in \mathbb{Z}$, where we do not need to specify the re-indexing $(\pm i, \pm j) \mapsto k$.

The size $(w, h)$ of the MLA is chosen such that the projection of the grid, after rotation $\alpha$ and tilt $(\beta, \gamma)$, covers the full sensor of size $(s_x, s_y)$, i.e. it can be calculated via
\begin{equation}\label{eq:mla-size}
	\myvec{R}_{x, \gamma}\,\myvec{R}_{y, \beta}\,\myvec{R}_{z, \alpha} \COLVEC{w}{h}{0} = \COLVEC{s_x}{s_y}{z} \,,
\end{equation}
where $z$ is arbitrary.

The perspective projection of the ML centers \eqref{eq:ml-centers} from the center $(0, 0, 0)^\TT$ of the exit pupil onto the sensor is given by
\begin{equation}
	\myvec{c}_{\pm i \pm j}^\textrm{p} = \lambda_{\pm i \pm j} \myvec{c}_{\pm i \pm j}
\end{equation}
with scaling factor $\lambda_{\pm i \pm j}$ such that
\begin{equation}
	\left( \myvec{c}_{\pm i \pm j}^\textrm{p} \right)_{\!z} = \left( \lambda_{\pm i \pm j} \myvec{c}_{\pm i \pm j} \right)_{\!z} =  - F - f \,,
\end{equation}
where $f$ denotes the ideal ML focal length. Therefore, using \eqref{eq:ml-centers}, we find
\begin{equation}\label{eq:lambda-factor}
	\lambda_{\pm i \pm j} = \frac{-F -f}{\left( \myvec{R}_{x, \gamma}\,\myvec{R}_{y, \beta}\,\myvec{R}_{z, \alpha}\myvec{c}_{\pm i \pm j}^{\textrm{id}} \right)_{\!z} - F} \;.
\end{equation}

The orthogonally projected centers $\myvec{c}_{\pm i}^\textrm{o}$ are simply obtained from the $\myvec{c}_{\pm i \pm j}$ by setting their $z$-value to $(-F - f)$.

\subsection{MLA accuracy estimates}\label{sec:accuracy-estimates}
In order to simplify some of the parameters, consider the following estimates.
Assuming an ideal grid, $0 = \alpha = \beta = \gamma$, $\myvec{0} = \myvec{\epsilon}_{k}$, the focal length $f_{k}$ of a ML has to be accurate within 
\begin{equation}\label{eq:f-acc}
\Delta_f < p f / d
\end{equation} in order for the disk of confusion to lie within a pixel \cite{Ng2005} with pixel pitch $p$. Deviations from this constraint will lead to blur in the decoded image which cannot be compensated. Following the same argument, the rotation $\alpha$ and tilt $(\beta, \gamma)$  have to be constrained such that the maximum change in distance $\Delta_z$ to the sensor fulfills the same restriction. To estimate this, we use the point $(w/2, h/2, 0)^\TT$ and rotate and tilt it using $\myvec{R}_{x, \gamma}\,\myvec{R}_{y, \beta}\,\myvec{R}_{z, \alpha}$. The resulting $z$-component then yields the maximum change of distance of the MLA to the sensor. Using \eqref{eq:mla-size}, we find
\begin{align}
	\Delta_z = s_x \big( \tan \beta / \cos \gamma \big) + s_y \tan \gamma < p f / d \,.
\end{align}
Note that the result does not depend on the rotation $\alpha$. To obtain a common upper bound $\Delta_\delta$ for the accuracies of the tilt angles, we perform a Taylor series expansion in $\beta, \gamma = 0$,
\begin{align}
\Delta_z \approx s_x \left(\beta + {\beta^3}/{3}  + {\beta \gamma^2}/{2} \right) + s_y \left( \gamma + {\gamma^3}/{3} \right) \,,
\end{align}
set $\beta = \gamma \equiv \Delta_\delta$ and solve
\begin{equation}\label{eq:max-tilt-accuracy}
	\Delta_z \approx \Delta_\delta (s_x + s_y) + \Delta_\delta^3(5s_x / 6 + s_y / 3) < p f / d \,.
\end{equation}

Additionally, the tilt introduces geometric distortion in the perspectively projected grid. That is, the ideally regular grid $\left\{\myvec{c}_{k} \right\}_{k}$ with constant grid spacing $d$ will be projected onto an irregular grid $\left\{\myvec{c}_{k}^\textrm{p} \right\}_k$ with a local grid spacing 
\begin{align}
d_{x, \pm i \pm j} &= \big\lVert \myvec{c}_{\pm i \pm j}^\textrm{p} - \myvec{c}_{\pm (i - 1)\pm j}^\textrm{p} \big\rVert \,, \notag \\
d_{y, \pm i \pm j} &= \sqrt{3}\big\lVert \myvec{c}_{\pm i \pm j}^\textrm{p} - \myvec{c}_{\pm i \pm (j - 1)}^\textrm{p} \big\rVert / 2 \,.
\end{align}
The maximum difference in local grid spacing is given using the largest ML indices $i_\textrm{max} = \lceil s_x/ 2d_x \rceil, j_\textrm{max} = \lceil s_y/ 2d_y \rceil$ by
\begin{align}\label{eq:dist-acc}
	\Delta_{d, x, \text{max}} &= \big\vert d_{x, i_\textrm{max} j_\textrm{max}} - d_{x, -i_\textrm{max} -j_\textrm{max}}\big\vert  \,, \notag \\
	\Delta_{d, y, \text{max}} &= \big\vert  d_{y, i_\textrm{max} j_\textrm{max}} - d_{y, -i_\textrm{max} -j_\textrm{max}}\big\vert   \,.
\end{align}
This formula can be used to estimate whether the tilt (if within the constraint \eqref{eq:max-tilt-accuracy}) is detectable in the ML image, as we exemplarily examine in the next section in the case of the Lytro Illum camera. For all mathematical and numerical estimates made above, we provide a simple Python script which helps checking the estimates in a given application.

\subsection{Reference data parameter choice}
In the remainder, all parameters are chosen according to a Lytro Illum light field camera. That is, we use a sensor of size $7728 \times 5368\,\si{px}$ with a pixel pitch of $p = \SI{1.4}{\micro\meter}$, a resolution of $10\,\si{bit}$ and a gamma factor of $0.4$. The MLs have an approximate diameter of $d = \SI{20}{\micro\meter}$ and a fixed f-number of $f/2$, hence an ideal focal length of $f = \SI{40}{\micro\meter}$. The MLs are arranged in a hexagonal grid with an estimated grid noise standard deviation of $\SI{0.1}{\percent}$ of the ML diameter, i.e.\ $\sigma_\textrm{g} = \SI{0.0143}{px}$. Unfortunately, it was not possible to find manufacturer specifications on the grid spacing accuracies so they had to be estimated. Of course, they can be adapted in the simulation when known in a specific application. The main lens of the Lytro Illum camera is a zoom lens with a focal length of \SI{30}{\milli\meter} to \SI{250}{\milli\meter}. The Lytro Illum camera provides a set of 33 different white images, taken at 10 different zoom settings and different focus settings. In order to be able to compare the synthetic results to actual white images, we choose four main focal lengths for which a corresponding white image is provided by the camera. In particular, we choose focal lengths $F$ of \SI{30}{\milli\meter}, \SI{47}{\milli\meter}, \SI{117}{\milli\meter}, and \SI{249}{\milli\meter}. For every white image, we simulate three different aperture settings, ranging from no mechanical vignetting to strong vignetting, where the object-side aperture is chosen such that the resulting vignetting effect is visually comparable with the Lytro white image of the corresponding zoom setting and a focus setting showing the strongest vignetting.

Following \eqref{eq:f-acc}, the ML focal lengths have to be accurate within $\Delta_f < \SI{2.8}{\micro\meter}$. Furthermore, according to \eqref{eq:max-tilt-accuracy}, assuming ideal ML focal lengths, the tilt has to be accurate within $\Delta_\delta < 0.0088^\circ$. The accuracy will have to be even higher in order for the combined $\Delta_{f} + \Delta_z$ to fulfill the constraint. For the maximum geometrical distortion within these constraints, following \eqref{eq:dist-acc}, we find $\Delta_{d, x, \textrm{max}} \approx \Delta_{d, y, \textrm{max}} = \SI{0.0022}{px}$ assuming a \SI{30}{\milli\meter} main lens for which the scaling factor $\lambda$ and distortion effects are the largest. The geometric distortion hence is negligibly small. Therefore, all images are synthesized with zero tilt, $\beta = 0 = \gamma$. When, in a given application, the geometric distortion is non-negligible, it can be included in the simulation and has to be estimated and corrected before estimating a regular ML grid from the WI.
The remaining parameters, such as the grid rotation $\alpha$ and offset $\myvec{o}_\textrm{g}$, are varied to obtain a collection of different WIs in order to increase the statistical significance of the evaluation. 

\begin{figure}
	\centering
	\includegraphics{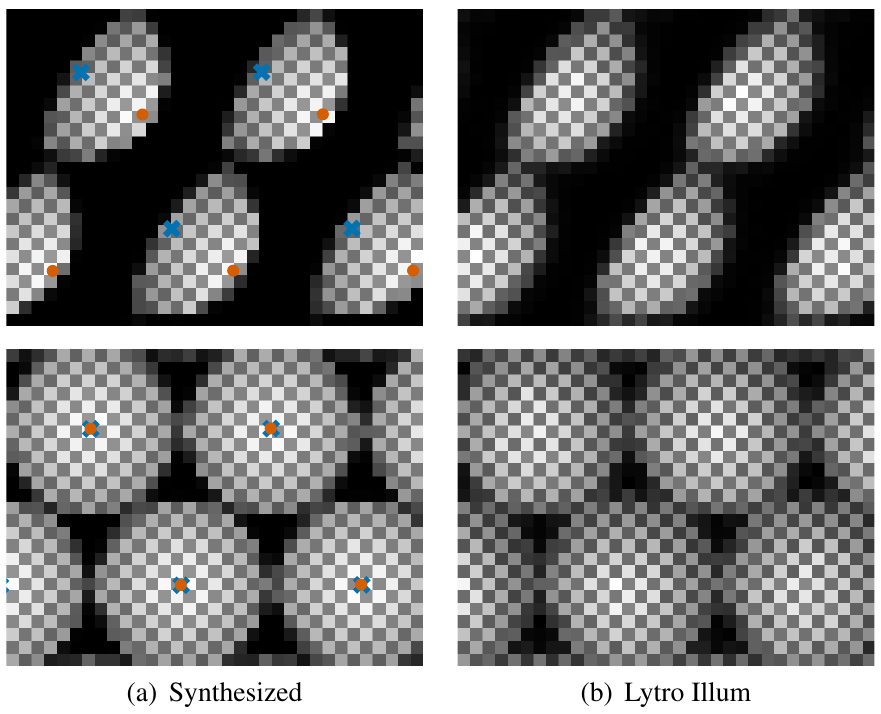}
	\caption{Comparison of a raw synthetic and Lytro Illum WI in the case of a \SI{30}{\milli\meter} main lens and strong mechanical vignetting. Crops are from the top left corner (top row) and center of the image (bottom row). The perspectively projected reference centers $\myvec{c}^\textrm{p}_k$ are depicted as blue crosses, the orthogonally projected $\myvec{c}^\textrm{o}_k$ as red dots.}
	\label{fig:white-images}
\end{figure}

We ray-trace a total of 240 white images. The synthetic WIs are then mosaiced using a Bayer pattern with color response according to a Lytro Illum camera. Furthermore, we add Gaussian image noise with standard deviation $\sigma_\textrm{n}$ of four different levels to the WIs to investigate the robustness of the grid estimation algorithms with respect to image noise. Hence, we evaluate a total of 960 different WIs.  A comparison of a synthesized and a Lytro Illum WI is shown in \FIG{fig:white-images}. The synthesized image incorporates all characteristics of the real one, in particular natural vignetting, which causes off-center brightest pixels, and mechanical vignetting resulting in the characteristic cat eye shape of the projected microlens images close to the sensor edges. 

%=======================================================================
\section{Microlens grid estimation}\label{sec:ml-detection}
%=======================================================================
The main purpose of pre-calibration is to estimate a regular grid approximating the perspectively projected ML centers $\myvec{c}_{k}^\textrm{p}$, which correspond to the  coordinates of the central rays of the target light field. In the further decoding pipeline, the estimated grid is used to align the lenslet image with the sensor and slice it to a 4D light field (see Section~\ref{sec:lf-decoding}).

Usually, the ML grid is estimated by detecting the ML centers from the corresponding WI \cite{Bok2017} and building a regular grid best approximating the detected centers \cite{Dansereau2013} or by directly estimating  regular grid from the WI \cite{Cho2013}. Challenges in the detection are versatile: on the one hand, the sheer amount of MLs in MLAs used in practice limits the algorithm's complexity. On the other hand, the geometry of the MLA is not trivial and usually slightly irregular. Furthermore, main lens and ML vignetting influences the (local) shape and brightness of the ML images, particularly of those that are close to the sensor edge: while ML images near the sensor center are circular and brightest in the center, ML images closer to the sensor edge are cat-eye-shaped and show an off-center brightest pixel (see \FIG{fig:white-images}). There are mainly two methods proposed in the literature: \CHO\ \cite{Cho2013} first compensate the rotation using an estimate obtained from the Fourier transform of the WI. In the spatial domain, they perform a grayscale erosion and clustering of the demosaiced WI. To estimate the ML centers, they use a parabolic least squares (LS) regression of the clustered MLs. In the decoding pipeline by \DANS\ \cite{Dansereau2013}, implemented as the de-facto light field decoding standard in the \MLTB, the raw WI is convolved with a disk kernel. The ML centers are then estimated by finding the local maxima in the filtered image. This does not result in subpixel precision. However, in the succeeding pre-calibration, the grid parameters are estimated with subpixel precision. Furthermore, Hog et al.\ \cite{Hog2017} present a Fourier-based estimation algorithm. But since it does not yield results that are significantly different from those of \DANS\ \cite{Dansereau2013}, we do not re-evaluate it here. For completeness, we will also evaluate the ML center detection used by Bok et al.\ \cite{Bok2017}, which is not presented in their paper but implemented in the corresponding \textsc{Matlab} reference implementation.

None of the mentioned algorithms consider vignetting effects. Taking into account the natural and mechanical vignetting by estimating the ML grid parameters and coordinates in the spatial domain of the WI is extremely challenging. Local circle search algorithms have been proposed \cite{Mignard2018} which show good performance in the image's central regions and in cases of strong vignetting close to the sensor edge, but mediocre performance in cases of only slight mechanical vignetting. Lytro supposedly uses a similar local arc fitting to account for mechanical vignetting \cite{Liang2016}, but, since their software is closed-source, this is speculative.

%We argue, that vignetting has an impact on the accuracy of the estimated grid parameters and hence on the light field decoding and calibration. For this, consider the following accuracy requirements.

\subsection{Grid estimation accuracy requirements}\label{sec:det-acc-req}

\begin{figure}[t]
	\centering
	\subfloat[]{\includegraphics{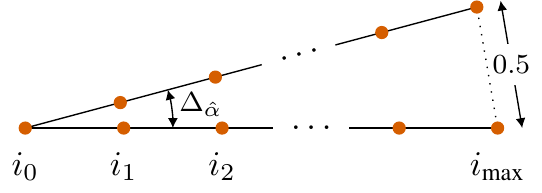}}
	\hfil
	\subfloat[]{\includegraphics{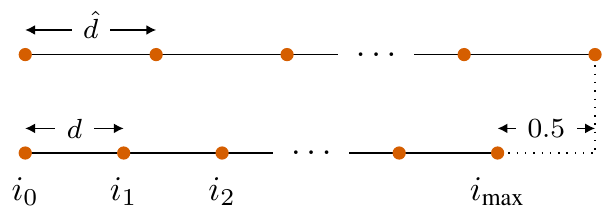}}
	\caption{Exaggerated sketch for the estimated spacing and rotation accuracy requirements. (a) Grid rotation accuracy. (b) Grid spacing accuracy.}
	\label{fig:grid-estimation-accuracy}
\end{figure}

The grid parameters (grid spacing, rotation, and offset) have to be estimated with very high accuracy. Assuming that the maximum deviation from a grid point of the estimated regular grid to a true grid point may not exceed $\SI{0.5}{px}$, we can estimate upper bounds of the individual accuracy requirements (compare \FIG{fig:grid-estimation-accuracy}).
Assuming that the grid offset is estimated perfectly and the grid matches the real grid at the sensor center, in order for the grid points furthest from the center to be within $\SI{0.5}{px}$ of the true grid centers, the accuracy $\Delta_{ \hat{d}}$ of the estimated grid spacing $\hat{d}$ has to be within the upper bound estimate
\begin{equation}
	\lvert \Delta_{\hat{d}} \rvert < \frac{0.5}{l_\textrm{max}} = \SI{0.0018}{px}
\end{equation}
in the case of the Lytro Illum camera. Here, 
\begin{align}
	l_\textrm{max} = \textrm{max}\!\left\{i_\textrm{max}, j_\textrm{max}\right\} = \textrm{max}\big\{\lceil 2s_x / d_x \rceil, \lceil 2s_y / d_y \rceil\big\}
\end{align}
is determined by the longer side of the sensor.
Following a similar argument, the accuracy $\Delta_{ \hat{\alpha}}$ of the estimated grid rotation $\hat{\alpha}$ has to satisfy
\begin{equation}
	\sin \lvert \Delta_{ \hat{\alpha}}\rvert < \frac{0.5 \cdot p}{ i_\textrm{max}\cdot d}
	\implies
	\lvert \Delta_{ \hat{\alpha}} \rvert < \arcsin \frac{0.5\cdot p}{ i_\textrm{max}\cdot d} = 0.0074^\circ \,.
\end{equation}
Finally, the grid offset leads to a global shift of the estimated grid, and should hence at least be accurate within
\begin{equation}
		\lvert \Delta_{\hat{o}} \rvert < \SI{0.5}{px} \,.
\end{equation}

\subsection{Proposed pre-calibration}\label{sec:pre-calibration}

\begin{figure}
	\centering
	\includegraphics{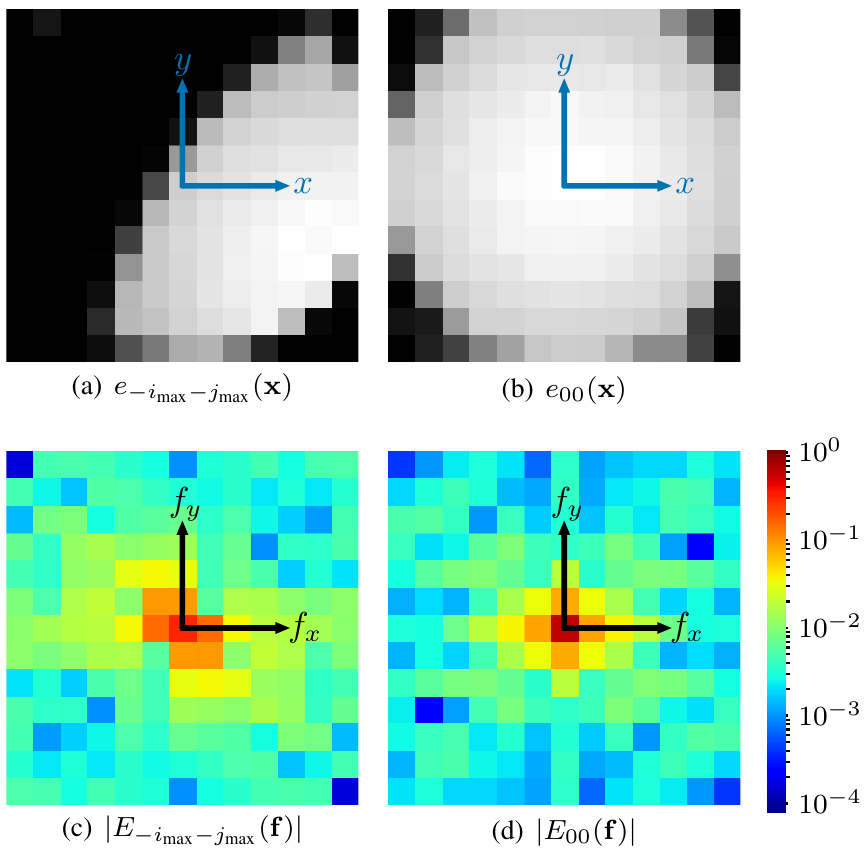}
	\caption{Comparison of central and upper left texel cropped from \FIG{fig:white-images} in the spatial domain (demosaiced, upper row) and in the Fourier domain (bottom row, normalized).}
	\label{fig:texel}
\end{figure}
The accuracy estimates made above pose challenging requirements on the estimation algorithms, in particular on the estimation of the grid spacing.
We propose a novel algorithm which operates in the Fourier domain to estimate the grid spacing and rotation, and in the spatial domain to estimate the grid offset. The estimation takes into account the natural and mechanical vignetting present in the white images. 

\subsubsection{Grid rotation and spacing estimation}
A white image can be interpreted as an (approximately) regular structure: ignoring natural and mechanical vignetting, a white image is made up of a regular texel $e(\myvec{x})$ which is arranged in a grid spanned by the vectors $\myvec{b}_1$ and $\myvec{b}_2$. For example, $\myvec{b}_1 = (0, 2r), \myvec{b}_2 = (\sqrt{3}\cdot r, r)$ for a per\-fect hexagonal grid with a ML radius of $r = d/2$. Using 2D convolution (denoted by $\CONV$), this results in the (continuous) white image
\begin{align}
g_{\mathrm{ideal}}(\myvec{x}) 
	&= e(\myvec{x}) \CONV \!\sum_{i, j \in \mathbb{Z}} \delta(\myvec{x} - i\,\myvec{b}_1 - j\,\myvec{b}_2) \\[-3mm]
	&\vTZz \notag \\
G_{\mathrm{ideal}}(\myvec{f}) 
	&\propto E(\myvec{f}) \cdot \sum_{i. j \in \mathbb{Z}} \delta(\myvec{f} - i\,\myvec{f}_1 - j\,\myvec{f}_2) \,,
\end{align}
where $E(\myvec{f})$ denotes the Fourier transform of $e(\myvec{x})$. For the frequency basis vectors $\myvec{f}_k = (f_{k, x}, f_{k, y})$, it holds \cite{Puente2016a}
\begin{align}\label{eq:basis-freq-to-spatial}
\begin{pmatrix}
b_{1, x} & b_{2, x} \\
b_{1, y} & b_{2, y}
\end{pmatrix} 
= 
\begin{pmatrix}
f_{1, x} & f_{1, y} \\
f_{2, x} & f_{2, y}
\end{pmatrix}^{-1} \,,
\end{align}
where $b_{k,x}, b_{k,y}$ are the components of the vectors $\myvec{b}_{k}$. Ideally, we can estimate the grid spacing and rotation by detecting the peaks corresponding to $\myvec{f}_{1}, \myvec{f}_{2}$ (and their multiples) in the absolute value of the Fourier transform $E(\myvec{f})$.

\begin{figure}
	\centering
	\includegraphics{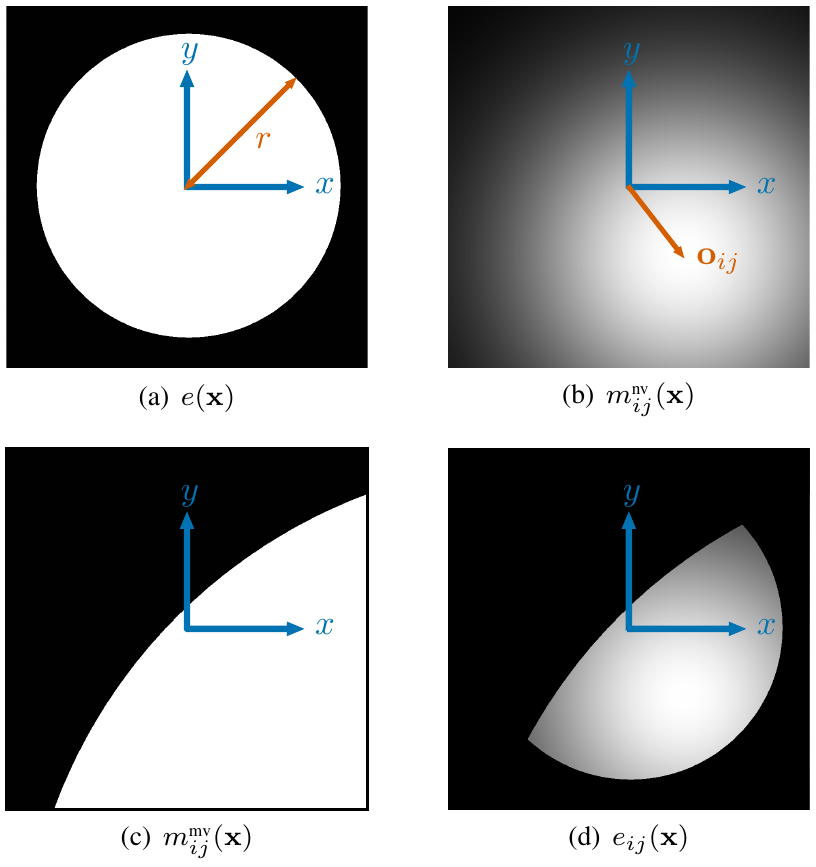}
	\caption{Schematic drawing of the continuous local texel model according to \eqref{eq:texel-model} in the case of strong mechanical vignetting}
	\label{fig:texel-model}
\end{figure}

Now, introducing natural and mechanical vignetting will not change the grid vectors but instead modulate the (now local) texel: in the image center, the texel will not be altered, but deviating from the center, the brightness distribution of the texel will change due to natural vignetting. Furthermore, due to mechanical vignetting, some pixels of the texel will be blocked. Therefore, the texel $e_{ij}(\myvec{x})$ is different at every grid position $(i, j)$ and we write
\begin{equation}\label{eq:grid-spatial}
g(\myvec{x}) = \sum_{i, j \in \mathbb{Z}} e_{ij}(\myvec{x})\CONV  \delta(\myvec{x} - i\,\myvec{b}_1 - j\,\myvec{b}_2) \,.
\end{equation}
As an example, two local (discrete) texels are shown in \FIG{fig:texel}. 
We calculate the Fourier transform of \eqref{eq:grid-spatial} to
\begin{align}\label{eq:wi-local-fourier}
G(\myvec{f}) = \sum_{i, j \in \mathbb{Z}} E_{ij}(\myvec{f}) \cdot  \ee^{-2\pi \ii \,(i \myvec{b}_1 + j \myvec{b}_2)\cdot \myvec{f} } \,,
\end{align}
where $E_{ij}(\myvec{f})$ denotes the Fourier transform of $e_{ij}(\myvec{x})$. Since every texel is different, \eqref{eq:wi-local-fourier} cannot directly be written as a Dirac comb like in the ideal case. 
We now model the local texels as
\begin{equation}\label{eq:texel-model}
	e_{ij}(\myvec{x}) = e(\myvec{x}) \cdot m^{\textrm{nv}}_{ij}(\myvec{x}) \cdot m^{\textrm{mv}}_{ij}(\myvec{x})
\end{equation}
where $e(\myvec{x})$ is a binary circular mask with ML radius $r$, $m^{\textrm{nv}}_{ij}(\myvec{x})$ is the modulation due to natural vignetting, whose shape we do not have to specify more explicitly but could for example model as a wide Gaussian bell, and $m^{\textrm{mv}}_{ij}(\myvec{x})$ describes the modulation due to mechanical vignetting, which can be modeled again as a binary circular mask of large radius with non-zero center. In more detail, the natural vignetting can be written as
\begin{equation}\label{eq:texel-model-modulation}
	m^{\textrm{nv}}_{ij}(\myvec{x}) = m^{\textrm{nv}}_{00}(\myvec{x} - \myvec{o}_{ij})
\end{equation}
where $\myvec{o}_{ij}$ is the distance of the perspectively projected to the orthogonally projected ML center on the sensor plane, since natural vignetting causes the brightest pixel to be at the orthogonally projected center, but the modulation shape does not change otherwise. A schematic drawing of the local texel model is shown in \FIG{fig:texel-model}. We calculate the Fourier transform of $e_{ij}(\myvec{x})$ using \eqref{eq:texel-model} and \eqref{eq:texel-model-modulation}:
\begin{align}
	E_{ij}(\myvec{f}) 
	&= E(\myvec{f}) \CONV M^{\textrm{nv}}_{ij}(\myvec{f}) \CONV M^{\textrm{mv}}_{ij}(\myvec{f}) \notag \\
	&= E(\myvec{f}) \CONV M^{\textrm{nv}}_{00}(\myvec{f})\cdot \ee^{-2\pi \ii \,\myvec{o}_{ij} \cdot \myvec{f}} \CONV M^{\textrm{mv}}_{ij}(\myvec{f}) \,,
\end{align}
where both the Fourier transform $E(\myvec{f})$ and $M^{\textrm{mv}}_{ij}(\myvec{f})$  are Airy discs of different widths (and phase).

We observed that natural vignetting will cause the periodic peaks in the Fourier transform of the WI to shift. This is likely due to the underlying periodic structure of the modulation itself which manifests itself in the phase factor: the natural vignetting of the overall WI can be seen as a regular texture which is arranged in the hexagonal grid of the orthogonally projected ML centers (instead of the perspectively projected ones). To eliminate this effect, we perform a strong gamma compression to effectively eliminate the modulation $m^{\textrm{nv}}_{00}(\myvec{x})$. That is, we calculate
\begin{align}
\tilde{g}_\gamma(\myvec{x}) 
	&= g^\gamma(\myvec{x})  \notag \\
	&\approx \sum_{i, j \in \mathbb{Z}} e(\myvec{x})\cdot  m^{\textrm{mv}}_{ij}(\myvec{x})   \CONV  \delta(\myvec{x} - i\,\myvec{b}_1 - j\,\myvec{b}_2) \\
	&\vTZz \notag \\
\tilde{G}_\gamma(\myvec{f}) 
	&\approx \sum_{i, j \in \mathbb{Z}} E(\myvec{f}) \CONV M^{\textrm{mv}}_{ij}(\myvec{f}) \cdot  \ee^{-2\pi \ii \,(i \myvec{b}_1 + j \myvec{b}_2)\cdot \myvec{f} } \,,
\end{align}
for $\gamma \ll 1$ such that 
\begin{equation}
	\left( m^{\textrm{nv}}_{00}(\myvec{x}) \right)^\gamma \approx 1 \TZz \delta(\myvec{f}) \,.
\end{equation} Note that since $e(\myvec{x})$ and $m^{\textrm{mv}}_{ij}(\myvec{x})$ are binary, they are unchanged from the gamma compression. This gamma compression has the additional advantage that the algorithm can equivalently operate on the raw, mosaiced WI, since the compression will push all gray values to one, mitigating the possible effects introduced by demosaicing algorithms, which are particularly severe in the case of microlens images \cite{David2017}. For the evaluation, we will hence apply our method to the raw WIs.

In practice, the texels will deviate from our texel model \eqref{eq:texel-model}, in particular, the $e(\myvec{x})$ and $m^{\textrm{mv}}_{ij}(\myvec{x})$ will not be exactly binary. The gamma compression then would have no effect as all non-zero pixels would be mapped to one. To make sure that the texel model \eqref{eq:texel-model} is acceptable, we perform contrast stretching prior to gamma compression. That is, for some $q \in (0, 1)$,we linearly map the value range $[q, 0.99]$ to $[0, 1]$ and clip all values below $q$ to zero and above $0.99$ to $1$. The value of $q$ depends on the white image and the camera parameters. The resulting white image is denoted by $\tilde{g}_{\gamma, q}$. Furthermore, since the mechanical vignetting only has an effect on off-center texels $(i, j) > (I, J)$ for some $I, J \in \mathbb{Z}$, we perform windowing using a rotationally symmetric Gaussian window $w_{\sigma}(\myvec{x})$ prior to calculating the Fourier transform. This suppresses off-center texels which are distorting the ideal spectrum due to mechanical vignetting. The standard deviation $\sigma$ of the Gaussian window is chosen such that
\begin{align}
\tilde{g}_{\sigma, \gamma, q}(\myvec{x})
	&= w_{\sigma}(\myvec{x})\cdot \tilde{g}_{\gamma, q}(\myvec{x}) \\
	&\approx w_{\sigma}(\myvec{x})\cdot \bigg( e(\myvec{x}) \CONV \!\sum_{i, j \in \mathbb{Z}} \delta(\myvec{x} - i\,\myvec{b}_1 - j\,\myvec{b}_2) \bigg)\,. \notag
\end{align}
Hence, the Fourier transform of can be approximated as
\begin{align}\label{eq:fourier-final-grid}
\tilde{G}_{\sigma, \gamma, q}(\myvec{f})
&\approx W_{\!\rho}(\myvec{f}) \CONV \bigg( E(\myvec{f}) \cdot \sum_{i, j \in \mathbb{Z}} \delta(\myvec{f} - i\,\myvec{f}_1 - j\,\myvec{f}_2)\bigg) \notag\\
&= \sum_{i, j \in \mathbb{Z}} W_{\!\rho}(\myvec{f} - i\,\myvec{f}_1 - j\,\myvec{f}_2) \cdot E(i\,\myvec{f}_1 + j\,\myvec{f}_2) \,,
\end{align}
where the $\myvec{f}_i$ are given by \eqref{eq:basis-freq-to-spatial} and the Fourier transform $W_{\!\rho}(\myvec{f})$ of the Gaussian window is again a Gaussian window with standard deviation $\rho = 1/(2\pi \sigma)$. 
Therefore, the Fourier transform of $\tilde{G}_{\sigma, \gamma, q}(\myvec{f})$ is given by a sum of shifted Gaussians centered at linear combinations of the grid frequency vectors $\myvec{f}_i$. Since the standard deviation $\sigma$ of the window in the spatial domain is much larger then the grid spacing, the standard deviation $\rho$ of the window in the Fourier domain is much smaller then the grid frequency spacing. More specifically, in the case of the Lytro Illum camera, using a standard deviation of $\sigma = \SI{100}{px}$ and a hexagonal grid spacing of $\SI{15}{px}$, we find a standard deviation of $\rho = \SI{0.0016}{px^{-1}}$ and a smallest distance of frequency basis vectors of $d_\textrm{f} = \SI{0.0770}{px^{-1}}$. Therefore, the center of each Gaussian in \eqref{eq:fourier-final-grid} lies outside the $\SI{51}{\sigma}$ neighborhood of the closest neighboring Gaussian. The local maximum of each Gaussian is accordingly virtually undisturbed by neighboring ones. Hence, the peaks in $\hat{G}_{\sigma, \gamma, q}(\myvec{f})$ approximate well the linear combinations of the grid frequency spacing vectors $\myvec{f}_i$.

In the discrete case we apply a rotationally symmetric Hann window whose width is determined by the length of the smaller dimension of the white image to reduce spectral leakage.

By estimating the peaks in the spectrum $\hat{G}_{\sigma, \gamma, q}(\myvec{f})$ of the contrast-stretched, gamma-compressed and windowed white image, we can estimate the grid spacing of the underlying perspectively project microlens centers via \eqref{eq:basis-freq-to-spatial}.  We estimate these basis vectors $\myvec{f}_i$ in the Fourier domain by finding the local maxima in the magnitude of the Fourier-transformed WI that correspond to the first $n$ multiples of the frequency basis vectors $n\myvec{f}_i$. The number $n$ of detected maxima depends on the application. in the case of the Lytro Illum camera, we are able to find $n = 5$ values per frequency basis vector.
We use zero padding and a centroid calculation to estimate the (sub-pixel) coordinate of those frequency vectors, i.e.\ for every cluster $\mathcal{C}_{i, n}$ around a peak corresponding to $n\myvec{f}_i$ we calculate
\begin{equation}\label{eq:com}
\hat{\myvec{f}}_{i, n}
= \frac{\sum_{j \in \mathcal{C}_{i, n}} \myvec{f}_j \; \tilde{G}_{\sigma, \gamma, q}(\myvec{f}_j)}{\sum_{k \in \mathcal{C}_{i, n}} \tilde{G}_{\sigma, \gamma, q}(\myvec{f}_k)} \,.
\end{equation}

In total, we have introduced three hyperparameters ($\sigma, \gamma$ and $q$) which our proposed method depends on. In order to determine these parameters appropriately (i.e., such that the made approximations hold), we need a certain measure which does not depend on any prior knowledge (e.g.\ the ML centers or the underlying grid spacing). To this end, we propose the following:
from the estimated frequencies $\myvec{f}_{i, n}$, we calculate their distances
\begin{equation}
	\hat{d}_{i, n} = \norm{\hat{\myvec{f}}_{i, n+1} - \hat{\myvec{f}}_{i, n}}
\end{equation}
which, ideally, should be constant in the case of a regular grid for $i = 1, 2$, respectively. Hence, the estimated standard deviation $\hat{s}_i$ of the samples $\hat{d}_{i, n}$ should ideally be zero. We can therefore use this estimated variance in the Fourier grid spacing to optimize the hyperparameters $\gamma, q$, and $\sigma$: the optimal parameters should show a minimal variance. For a specific task, e.g.\ given a light field camera with a fixed prime lens, this minimization could be done manually. But for the Lytro camera, the white images corresponding to the different focal lengths show different features in terms of vignetting, grid spacing, and brightness distribution. Therefore, and to obtain an automated calibration process, we use differential evolution on a predefined search space of the hyperparameters to minimize the estimated standard deviation $\hat{s}_i$ of the Fourier grid spacing automatically. 

\begin{figure}
	\centering
	\includegraphics{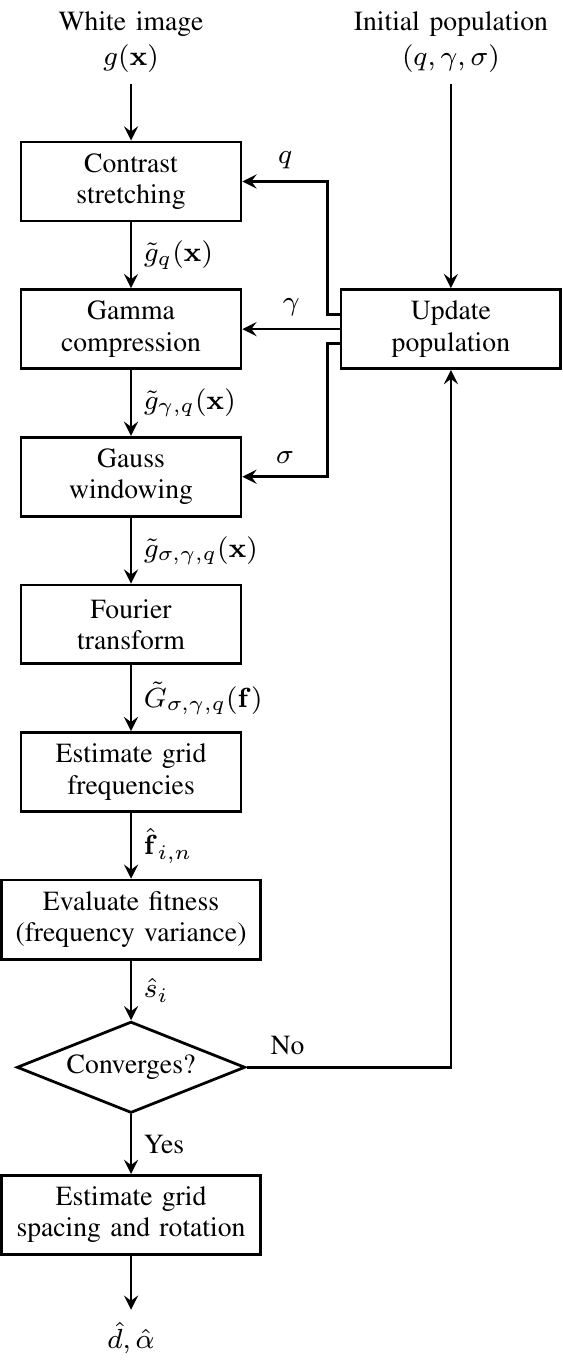}
	\caption{Flowchart of the proposed grid spacing and rotation estimation algorithm. The detailed steps for the minimization of $\hat{s}_i$ via differential evolution have been omitted for clarity.}
	\label{fig:flowchart-grid}
\end{figure}

Using the final estimated grid frequency vectors $\hat{\myvec{f}}_k$ with the corresponding estimated grid spacing vectors $\hat{\myvec{b}}_k$, we straightforwardly obtain estimates of the grid spacing and rotation of the perspectively projected ML grid. An overview of the proposed grid estimation is shown in \FIG{fig:flowchart-grid}.

\begin{figure}
	\centering
	\includegraphics{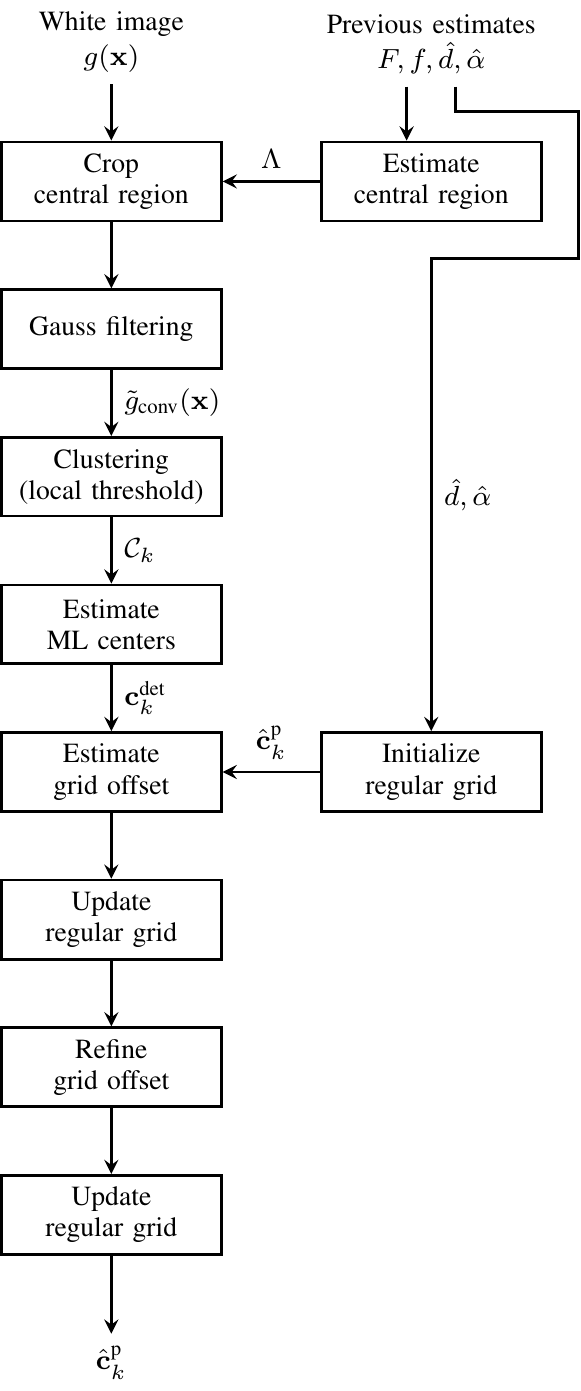}
	\caption{Flowchart of the proposed full grid estimation algorithm with offset refinement.}
	\label{fig:flowchart-offset}\vspace{-2mm}
\end{figure}

\subsubsection{Grid offset estimation}
Having estimated the grid rotation and spacing, we are left to estimate the overall grid offset, which is done in the spatial domain. \DANS\ \cite{Dansereau2013} estimate the offset by building an initial regular grid, using the previously estimated grid spacing and rotation, and measuring the median distance of the regular grid points to the previously detected ML centers.

Here, we propose a refinement of this method. First, we estimate the ML centers only in the central region of the image where the expected difference of perspectively to orthogonally projected centers is less then $\SI{0.5}{px}$. This increases the accuracy of the detection since the orthogonally projected centers are more easy to detect due to natural vignetting. That is, we are restricting the detection region, denoted by $\Lambda$, such that
\begin{equation}
	\norm{\myvec{c}_{\pm i_\textrm{max}0}^\textrm{p} - \myvec{c}_{\pm i_\textrm{max}0}^\textrm{o}} = d \cdot i_\textrm{max} (\lambda_{\pm i_\textrm{max}0} - 1) < 0.5 \,.
\end{equation}
Since $d_x > d_y$, this will suffice to fulfill the analogous constraint in the $y$-direction as well. We will use a rough estimate of the factor $\lambda \approx (F + f) /F$ obtained from \eqref{eq:lambda-factor}, given a rough estimate of the image distance $F$ and the ML focal length $f$, and the previously estimated MLA spacing $\hat{d}$. The restricted region $\Lambda$ will be comparatively small, depending on the factor $\lambda$ and hence on the main lens focal length, consisting of as few as $50 \times 50$ MLs in the case of a \SI{30}{\milli\meter} main lens. While this small region is not suited to estimate the grid spacing with high accuracy, estimating the overall offset can be done with much fewer measurements, i.e.\ fewer available ML centers.
Having restricted the detection region, we convolve the image with a Gaussian kernel to reduce image noise. In a second step, the image is clustered using local thresholding (by local Gaussian weighted mean with a block size of 17\,px), to find areas around local peaks, and a standard cluster labeling algorithm. Each cluster represents exactly one ML. 
%\newpage
Finally, we estimate the ML centers from the detected clusters. That is, for each detected cluster $\mathcal{C}_k$, we calculate the center of mass, analogously to the calculation \eqref{eq:com} in the Fourier domain,
\begin{equation}
\myvec{c}_k^{\textrm{det}}
= \frac{\sum_{m \in \mathcal{C}_k} \myvec{x}_m \; g_\textrm{conv}(\myvec{x}_m)}{\sum_{n \in \mathcal{C}_k} g_\textrm{conv}(\myvec{x}_n)} \,.
\end{equation}
To estimate the grid offset, analogously to \DANS\ \cite{Dansereau2013}, we first calculate the median distance of the initialized regular grid points $\hat{\myvec{c}}_k^{\textrm{p}}$ to the detected ML centers $\myvec{c}_k^{\textrm{det}}$.
Additionally, in a refinement step, we will then calculate a \emph{weighted} median distance of the updated regular grid point to detected ML centers, assigning a higher weight to those ML centers that are more central (using a symmetric Gaussian window). Since the detection inaccuracies due to natural vignetting will be smaller in the image center, this should yield a more reliable final result of the estimated grid offset. An overview of the proposed offset estimation method is depicted in \FIG{fig:flowchart-offset}.

Using the estimated grid spacing, rotation, and offset, we calculate a final, estimated, regular hexagonal grid $\left\{\hat{\myvec{c}}_{k}^{\textrm{p}} \right\}_{k}$ approximating the perspectively projected ML centers $\myvec{c}_{k}^{\textrm{p}}$. 

\subsection{Quality measures}
To quantitatively measure the performance of the grid estimation algorithms, we use the following quality measures. 
We measure the overall grid estimation accuracy $Q_\textrm{g}$ by calculating the root mean square distance of estimated to true grid points:
\begin{equation}
Q_\textrm{g} = \sqrt{\frac{1}{M} \sum_{k = 1}^{M} \lVert \hat{\myvec{c}}_k^{\textrm{p}} - \myvec{c}_k^\textrm{p}\rVert^2 } \,.
\end{equation}
Here, $M$ denotes the number of grid points in the estimated grid. When grid noise has been added to the ideal grid points, we will measure higher values of $Q_\textrm{g}$. Ideally, we would estimate the perfect regular grid that the grid points are derived from, i.e.\ we obtain
\begin{equation}
Q_\textrm{g, ideal} 
= \sqrt{\frac{1}{M} \sum_{k = 1}^{M} \epsilon_k^2 }
= \frac{\sigma_g}{\sqrt{M}} \sqrt{\sum_{k = 1}^{M} \left(\frac{\epsilon_k}{\sigma_\textrm{g}} \right)^2 }
=: \frac{\sigma_g}{\sqrt{M}} X \,
\end{equation}
where $\epsilon_k \sim \mathcal{N}(0, \sigma_\textrm{g})$ and hence $X$ is distributed according to the chi distribution with $M$ degrees of freedom. We find the expected value
\begin{align}\label{eq:q_g_ideal}
\mathbf{E}\big[Q_\textrm{g, ideal}\big]
= \olit{Q}_\textrm{g, ideal}
= \sigma_\textrm{g}\cdot \sqrt{\frac2M} \cdot \frac{\Gamma((M +1)/2)}{\Gamma(M/2)} \,,
\end{align}
where $\Gamma$ denotes the gamma function. Using the identity
\begin{equation}
\lim\limits_{n \to \infty}\frac{\Gamma(n + \gamma)}{\Gamma(n)n^\gamma} = 1 \,,\quad \textrm{for all } \gamma \in \mathbb{C}\,,
\end{equation}
with $n = M/2$ and $\gamma = 1/2$, we find the approximation
\begin{equation}
	\olit{Q}_\textrm{g, ideal} \approx \sigma_\textrm{g}\,.
\end{equation}
We will view this as the ideal mean grid estimation accuracy.
To gain further insight in the grid estimation performance, we calculate the mean absolute difference of estimated to true grid rotation ${\alpha}$ as well as the mean absolute difference of estimated to true grid spacing $d$,
\begin{equation}
Q_\textrm{s} = \lvert \hat{d} - d\rvert \;,\quad
Q_\textrm{r} = \lvert \hat{\alpha} - {\alpha}\rvert \,.
\end{equation}
Furthermore, we measure the runtime of each grid estimation algorithm. The evaluation was carried out on 8 cores of an AMD EPYC 7351 CPU@2.40\,GHz  utilizing multithreading where possible.

\begin{figure*}[th]
	\centering
	\includegraphics{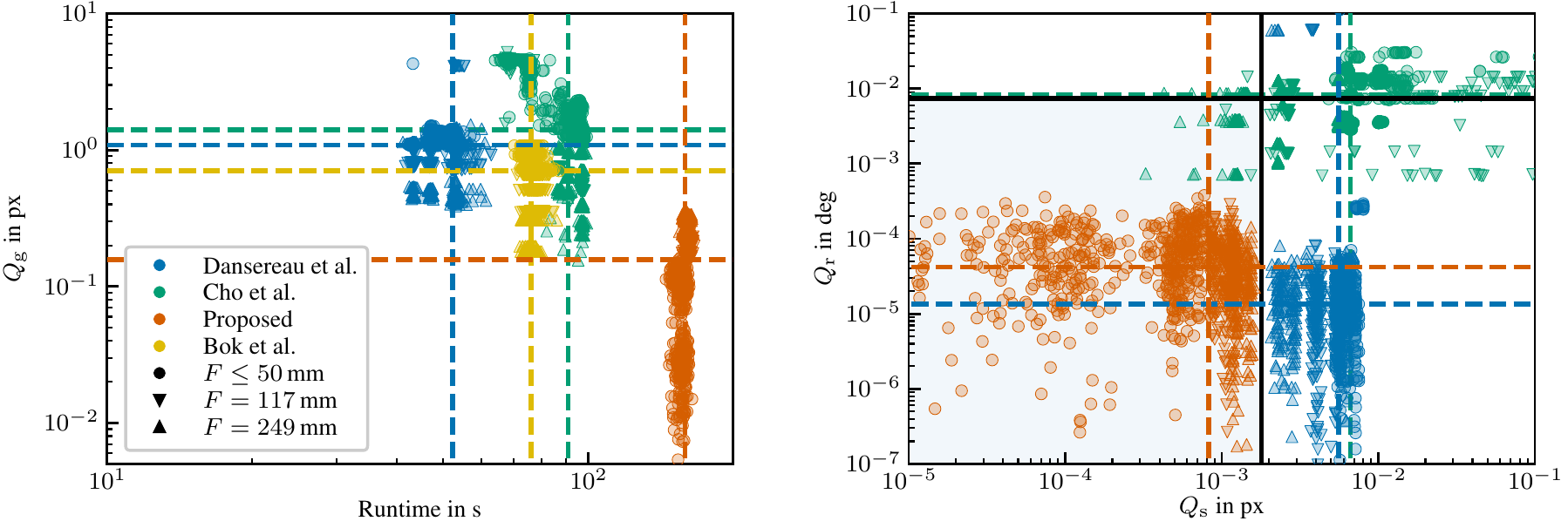}
	\caption{Performance comparison of the different ML grid estimation algorithms. The drawn colored lines show the median values of the corresponding datasets. The solid black lines denote the accuracy requirements as stated in Section~\ref{sec:det-acc-req}.}
	\label{fig:grid-est-comp}
\end{figure*} 

\subsection{Results}\label{sec:detection-results}
For all 960 WIs, a regular grid is estimated with the different estimation algorithms and the overall grid accuracy $Q_\textrm{g}$ as well as the spacing and rotation accuracies $Q_\textrm{s}, Q_\textrm{r}$ are calculated using the ground truth ML centers and grid parameters. A detailed comparison of the overall grid estimation performances is shown in \FIG{fig:grid-est-comp}. Note that, since the calibration by Bok at al.\ \cite{Bok2017} does not utilize a regular grid but the individually detected centers, the grid spacing and rotation accuracies $Q_\textrm{s}, Q_\textrm{r}$ cannot be specified in this case.
\begin{figure}[t]
	\centering
	\includegraphics{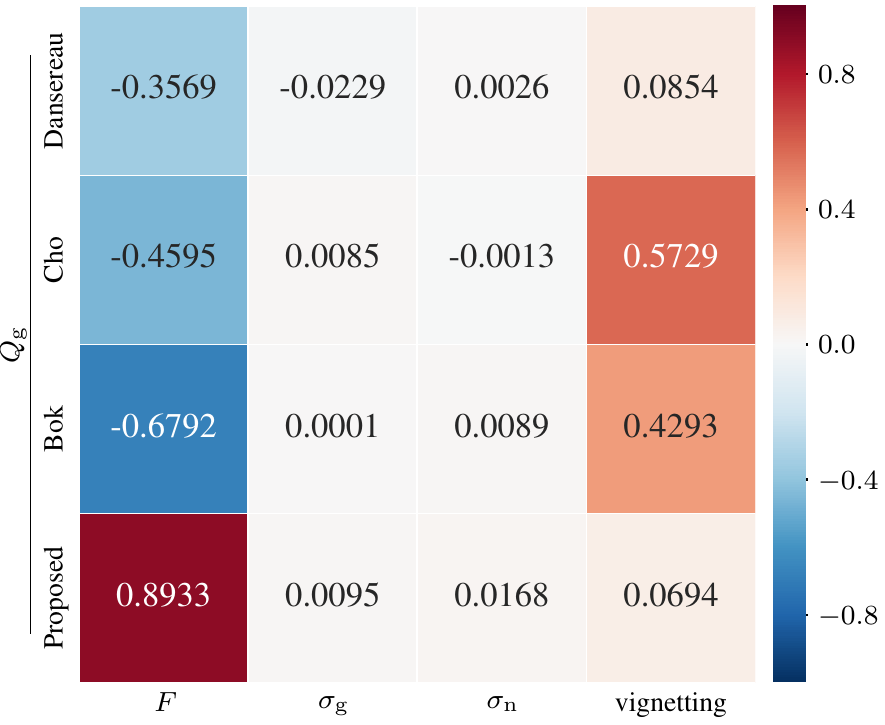}
	\caption{Pearson correlation of the grid estimation accuracy $Q_\textrm{g}$ with the image distance $F$, grid noise $\sigma_\textrm g$, image noise $\sigma_\textrm n$ and (mechanical) vignetting for the different estimation algorithms.}
	\label{fig:corr-total}\vspace{-5mm}
\end{figure}
We observe that only the proposed algorithm satisfies the accuracy requirements as stated in Section~\ref{sec:det-acc-req}. While the algorithm by \DANS\ performs very well in estimating the grid rotation, its spacing estimation is robust but limited in accuracy, in particular for short main lens focal lengths. The proposed method outperforms the others in the overall grid estimation accuracy. The improved accuracy stems from the more accurate grid spacing estimation while the performance regarding the rotation accuracy is slightly worse than the algorithm by \DANS\ \cite{Dansereau2013}. Still, the rotation estimation is performed with very high accuracy of about $10^{-4}\,\deg$. Conversely, the method proposed by \CHO\ does not yield a robust estimation of the grid spacing and rotation. In terms of the grid accuracy $Q_\textrm{g}$, we find that the proposed method yields results of about one order of magnitude better than the other methods while occasionally showing extreme results performing even two orders of magnitude better. We observe that the grid estimation proposed by Bok et al.\ \cite{Bok2017} slightly outperforms the one by \DANS\ \cite{Dansereau2013}.

Even though the proposed method has a longer runtime, with an average of about $\SI{160}{\second}$ per WI compared to an average of about $\SI{75}{\second}$ in the case of the method by Bok et al.\ and about $\SI{50}{\second}$ in the case of the method by \DANS\, we argue that this is still feasible, as the calibration usually only has to be executed once per camera. Furthermore, the runtime will be shorter in practice using a desktop PC with a clock speed higher than the used $\SI{2.4}{\giga\hertz}$.

\begin{table}
\centering
\caption{Mean grid estimation accuracy for the different algorithms for different image distances $F$ (in mm). All other quantities in pixel.}
\label{tab:grid-imagedist-comp}
\begin{tabular}{ccccccc}
	\toprule
	\multirow{3}{*}{\vspace{1mm}$F$} & \multirow{3}{*}{\vspace{1mm}$\sigma_\textrm{g}$} &	\multirow{3}{*}{\vspace{1mm}$\olit{Q}_\textrm{g, ideal}$} &		& $Q_\textrm{g}$ & 					  \\ \cmidrule{4-7}
	& 				&		& Dansereau & Cho &Bok			& Proposed 			\\
	\toprule
	30 	&     0\phantom{.0000} 	&    0\phantom{.0000} 	&   1.2850 &  2.4631 &  0.9724 &   \textbf{0.0865} \\
	&     0.0143 			&    0.0143 			&   	1.2855 &  2.6117 &  0.9723 &   \textbf{0.0881} \\[1mm]
	47 	&     0\phantom{.0000} 	&    0\phantom{.0000} 	&   1.1323 &  1.8162 &  0.7124 &   \textbf{0.0498} \\
	&     0.0143 			&    0.0143 				&   1.1075 &  1.6608 &  0.7126 &   \textbf{0.0561} \\[1mm]
	117 &     0\phantom{.0000} 	&    0\phantom{.0000} 	&   0.9418 &  2.8216 &  0.5056 &   \textbf{0.1973} \\
	&     0.0143 			&    0.0143 				&   0.9420 &  2.8486 &  0.5057 &   \textbf{0.1990} \\[1mm]
	249 &     0\phantom{.0000} 	&    0\phantom{.0000} 	&   0.8238 &  0.6398 &  0.4339 &   \textbf{0.2949} \\
		&     0.0143 			&    0.0143 			&   0.7613 &  0.6369 &  0.4340 &   \textbf{0.2913} \\
	\bottomrule
\end{tabular}
\end{table}

Investigating the results in more detail, as shown in \FIG{fig:corr-total}, we find that all methods are insusceptible to image and grid noise. While the method proposed by \CHO\ and the method by Bok et al.\ show a strong dependency on the mechanical vignetting present in the white image, the proposed method and the method by \DANS\ do not show such correlation. On the other hand, there seems to be a strong dependency on the image distance for all pre-calibration methods. While the methods by Bok et al.\, Cho et al., and \DANS\ perform increasingly better with larger image distances, the accuracy of the proposed method decreases. This is further analyzed in \TAB{tab:grid-imagedist-comp}. Since the scaling factor $\lambda$ converges to $1$ when the image distance increases, the influence of natural vignetting in the white image decreases. That is, with larger image distances, the orthogonally projected centers and the perspectively projected centers coincide. Hence, the methods relying on the local brightness distribution of every ML, such as the method by Bok et al.\ or \DANS, show an increase in accuracy. On the other hand, the proposed method shows extremely accurate estimates, close to the expected ideal mean accuracy $\olit{Q}_\textrm{g, ideal}$, in the case of a $\SI{30}{\milli\meter}$ and $\SI{47}{\milli\meter}$ main lens but a decreasing performance in the case of the $\SI{117}{\milli\meter}$ and the $\SI{249}{\milli\meter}$ lens. This is likely due to the characteristics of the mechanical vignetting. While for shorter main lens focal lengths, the mechanical vignetting only influences MLs very close to the sensor edge but with a sharp cutoff, the vignetting is more spread out in the case of a longer focal length. Therefore, the proposed algorithm is likely to use a smaller window size which decreases the estimation accuracy, since the effective resolution of the Fourier-transformed image is decreased. Still, the performance in those cases is better than the estimation accuracy reached by Bok et al.\, \DANS\ or Cho et al. Also, using a high-quality main lens for very long focal lengths should mitigate the effects of the mechanical vignetting and lead to higher estimation accuracies.

Overall, the proposed grid estimation algorithm outperforms the ones by \DANS\ \cite{Dansereau2013}, \CHO\ \cite{Cho2013}, and Bok et al.\ \cite{Bok2017}. As the method by \CHO\ could not provide reliable results, we exclude it from the remaining evaluation.
Next, we will evaluate how the increased grid estimation accuracy influences the light field decoding and calibration quality.

\begin{figure*}
	\centering
	\includegraphics{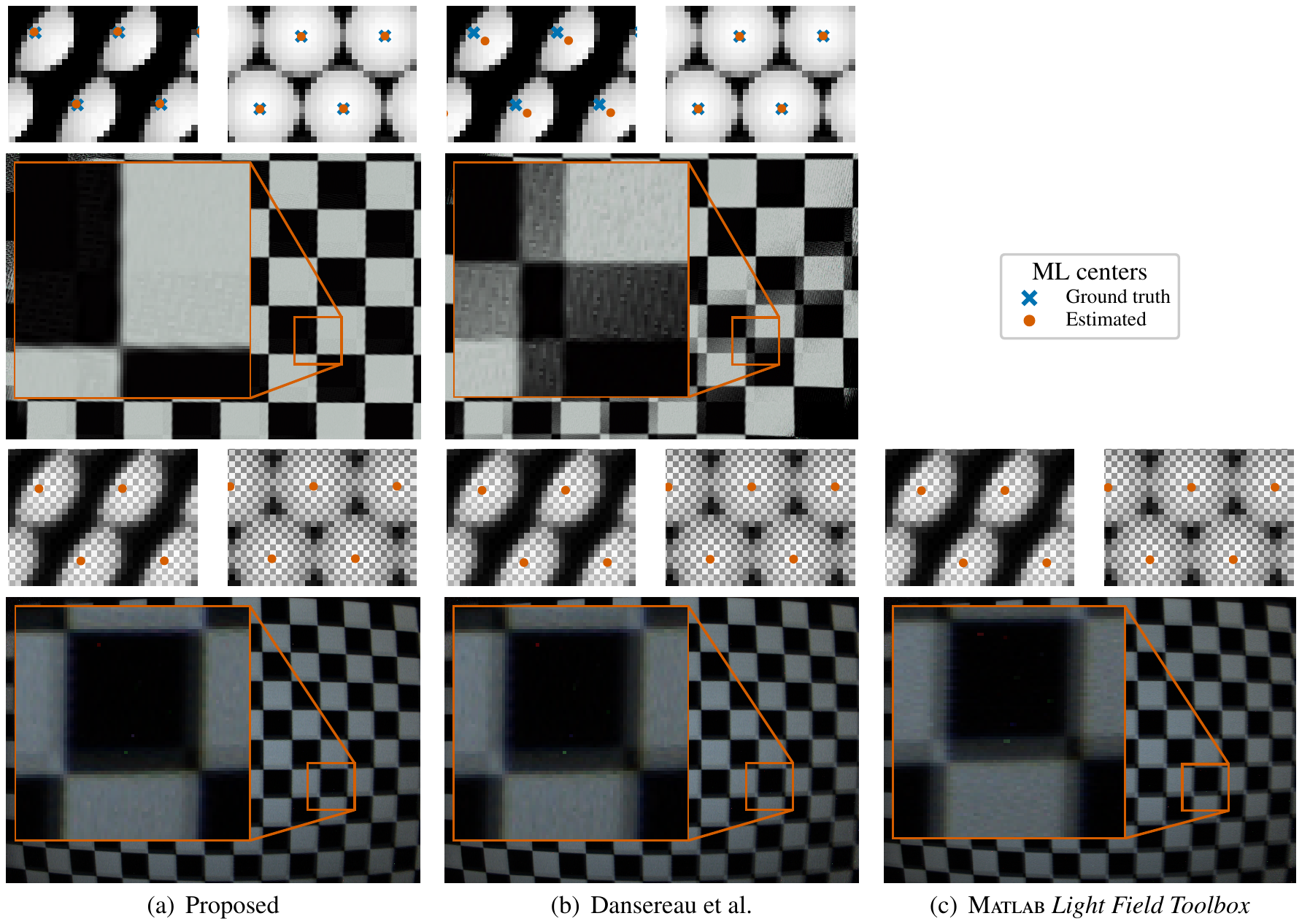}
	\caption{Comparison of peripheral subapertures decoded from the raw lenslet images and corresponding white image sections using different MLA grid estimation methods. Upper two rows: from synthetic raw images with known ground truth ML centers. Lower two rows: from images taken with a Lytro Illum camera.}
	\label{fig:lf-decoding-comp}
\end{figure*}

%=======================================================================
\section{Light field decoding}\label{sec:lf-decoding}
%=======================================================================

% Dansereau: chosen by JPEG-Pleno
As an example MLA-based application, we decode the raw sensor images of a Lytro Illum light field camera (and its ray tracer implementation) and investigate the decoding quality for the different ML centers and grid estimation algorithms. For the camera pre-calibration, an ideal grid is estimated as described in Section \ref{sec:ml-detection} for each of the 34 WIs that are provided by the camera. Each white image is taken at a different main lens focal length and focus distance.
The used decoding pipeline mostly follows \DANS\ \cite{Dansereau2013} and consists of the following: black and white level correction as specified in the raw image's meta data file, devignetting and color correction, using the white image corresponding to the zoom and focus settings of the raw image, and demosaicing, here using the method by Malvar et al.\ \cite{Malvar2004}. Using the estimated grid parameters, the sensor image is aligned with the grid by rotating, scaling, and translating the image, such that the ideal grid point coordinates fall on pixel centers and rotation is compensated. Finally, the sensor images can be sliced to a light field and resampled to a rectangular grid. For the last step, we use a gradient-guided interpolation technique, deviating from the 1D- or Delaunay-based interpolations that are used in the \MLTB: depending on the gradients of the image in $x$-, $y$- or $xy$-direction, we either perform 1D vertical, horizontal, or 2D bilinear interpolation to resample the light field.

The proposed decoding pipeline is included in a new Python framework that we release under an open-source license for the scientific community \cite{git}. All shown results, if not explicitly stated otherwise, are obtained using the proposed framework for comparability. All presented images were taken with a Lytro Illum camera, or its ray tracer implementation, set to a \SI{30}{\milli\meter} main lens focused at infinity.

\subsection{Results}

Two peripheral subapertures of two decoded example light fields, one synthesized using the proposed camera model and ray tracer implementation, and one taken with a Lytro Illum camera, for different MLA grid estimation algorithms together with their respective estimated ML centers and ideal grid point centers, are shown in \FIG{fig:lf-decoding-comp}. 

Similarly to the results of the previous section, we observe that the proposed ML grid estimation performs best: at the image center as well as the sensor edges, the proposed method approximates the ground truth ML centers with high accuracy while the method by \DANS\ slightly underestimates the grid spacing and hence the ML centers at the sensor edge are not correctly approximated by the regular grid (see top row of \FIG{fig:lf-decoding-comp}). A similar observation can be made for the real WI of a Lytro Illum camera. Even though the true ML centers are unknown, the method by \DANS\ again seems to slightly underestimate the grid spacing compared to the proposed method (see third row of \FIG{fig:lf-decoding-comp}). The results of our framework, using the method by \DANS, agrees with the results obtained with the \MLTB, validating our re-implementation of Dansereau's methods.

Regarding the decoding quality, we observe a slight improvement by using the proposed MLA estimation: in peripheral subapertures, ghosting artifacts appear less dominant (see last row of \FIG{fig:lf-decoding-comp}). The effect can be observed for both the synthetic as well as the real lenslet images. The effect in the synthesized images seems to be stronger, probably due to a stronger simulated mechanical vignetting.

Overall, we can conclude that the MLA grid estimation accuracy influences the light field decoding quality, even though the influence is subtle and mostly significant for peripheral subapertures. Nevertheless, even a slight improvement could make more subapertures available for further light field analysis, such as depth estimation and refocusing. In principle, the decoding quality could be further improved, e.g.\ by using an optimized demosaicing scheme \cite{David2017} or a different interpolation method for optimal hexagonal to rectangular resampling \cite{Cho2013}.

%=======================================================================
\section{Light field calibration}\label{sec:lf-calibration}
%=======================================================================

To further quantitatively evaluate the influence of the ML grid estimation accuracy, we perform a calibration of a Lytro Illum camera, again set to a focal length of \SI{30}{\milli\meter} and focused at infinity. For the calibration, we use two different methods that have been proposed in the literature and for which the source code is publicly available. Namely the calibration by \DANS\ \cite{Dansereau2013}, using corner features in the decoded light fields, and the calibration by Bok et al.\ \cite{Bok2017} which directly utilizes the raw lenslet images, using line features. Since the different methods rely on different features during the calibration, we use a different, optimized dataset for each. While the calibration by \DANS\ profits from many corners being present in the calibration images (and hence from smaller grid sizes), the method by Bok et al.\ performs better with larger grid sizes that show more line features in the raw lenslet images.

For the calibration using line features, we created a dataset containing \num{10} images of a checker grid with a baseline of \SI{15.57}{\milli\meter}. The calibration was performed using the \textsc{Matlab} code provided by Bok et al.\ \cite{Bok2017} which we have modified to be able to run the calibration with our previously estimated ML centers. We have performed the calibration using the unmodified code by Bok et al.\ as well as with the estimated ML grid points using the method by \DANS\ and our proposed method. Across all images and subapertures, the ray re-projection root mean square error (RMSE) as well as the projection RMSE is calculated.

For the calibration using corner features, we again created a dataset containing \num{10} images of a checker grid, now with a baseline of \SI{6.23}{\milli\meter}. The calibration was performed using the \textsc{Matlab} \textit{Light Field Toolbox} by Dansereau. We have slightly modified the code such that we could include the different MLA grid estimation results. Again, we calculate the ray re-projection RMSE across all images and subapertures.

\begin{figure*}
	\includegraphics{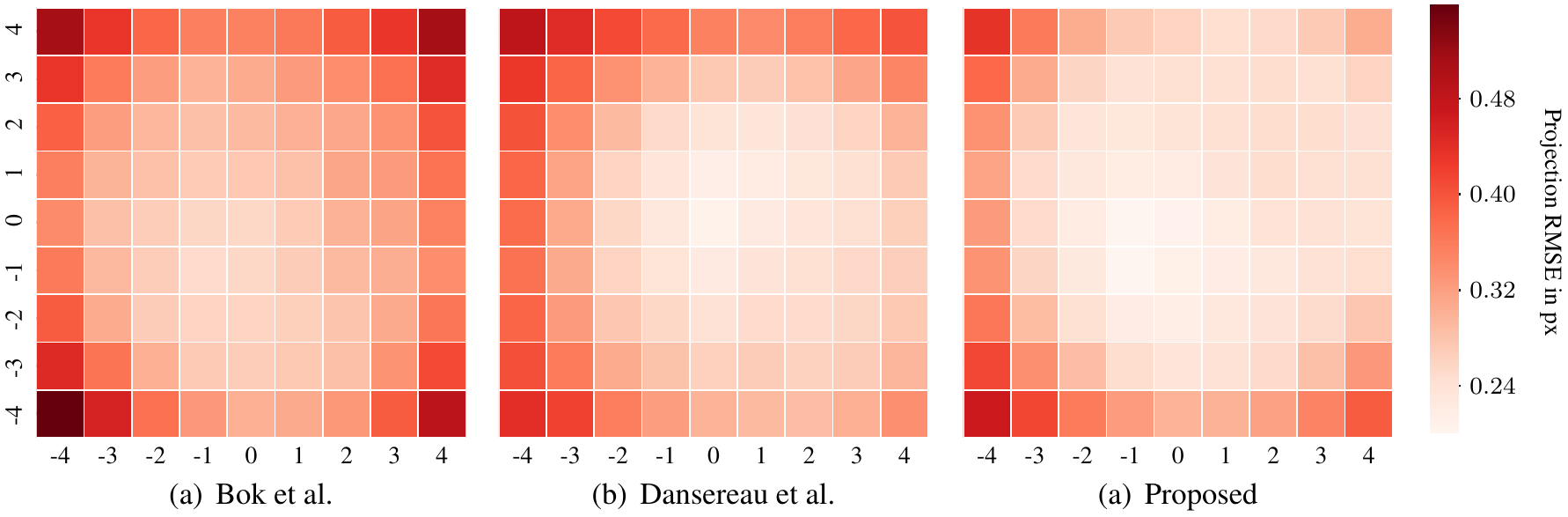}
	\caption{Projection RMSE across all calibration images for the different subaperture indices $(u, v)$ for different ML estimation algorithms in the case of calibration using line features for a Lytro Illum camera.}
	\label{fig:cal-results}
\end{figure*}
\subsection{Results}
The calibration results are shown in \TAB{tab:cal-results} and, in the case of the calibration using line features, in more detail in \FIG{fig:cal-results}. We observe that both calibration methods, the calibration using line features as well as the calibration using corner features, profit from the improved accuracy of the proposed MLA grid estimation algorithm. Depending on the used calibration, we are able to improve the overall ray reprojection RMSE by about \SI{0.02}{\milli\meter} to \SI{0.04}{\milli\meter} which is an improvement by about $\SI{15}{\percent}$ to $\SI{20}{\percent}$ compared to previous methods. Similarly to the results from Section~\ref{sec:lf-decoding}, the gain in accuracy is larger in peripheral subapertures (as shown in \FIG{fig:cal-results}). Again, this reflects the more accurate grid spacing estimation, which we observed in Section~\ref{sec:ml-detection}, and can make further light field analysis more robust and accurate.  

Additionally, in the case of the calibration using line features, we observe that the methods utilizing a regular grid to estimate the ML centers (namely the method by \DANS\ as well as the proposed method) result in a higher calibration accuracy. To some extent, this is surprising, as the results in Section~\ref{sec:detection-results} showed a slightly better ML center estimation performance in the case of the algorithm used by Bok et al.\ \cite{Bok2017}. However, note that in this instance, no regular grid approximating the centers in estimated.
This suggests two conclusions: first, the ML centers are more robustly estimated when approximated by a regular grid. When estimating a regular grid, a multitude of measurements are fused to obtain an accurate and robust result. Secondly, systematic, non-rigid deformations of the MLA are likely negligible (in the case of the used Lytro Illum camera). These irregularities should well be detected in the algorithm by Bok et al.\ which however shows a worse calibration result. This reinforces our decision to not include these deformations in our camera model.

\begin{table}
	\centering
	\caption{Root mean square errors across all subapertures of all calibration images for the different calibration and grid estimation methods for a Lytro Illum camera.}
	\begin{tabular}{llrr}
		\toprule
		Calibration & ML estimation & Ray reprojection& Projection\\
		method & method 	& RMSE in mm & RMSE in px \\
		\midrule
		Line& Bok et al.       &                 0.1409 &          0.4747 \\
		features & Dansereau et al. &                 0.1263 &          0.4208 \\
		{} & \textbf{Proposed}         &                 \textbf{0.1118} &          \textbf{0.3719} \\[2mm]
		Corner &  Dansereau et al. &                 0.2583 &          - \\
		features&  \textbf{Proposed}          &                 \textbf{0.2196} &          - \\
		
		\bottomrule
	\end{tabular}
	\label{tab:cal-results}
\end{table}

%\begin{figure*}
%	\includegraphics{figures/08-calibration-bok-rrpe}
%	\caption{Ray reprojection RMSE across all calibration images for the different subaperture indices for different ML estimation algorithms in the case of calibration using line features for a Lytro Illum camera.}
%	\label{fig:cal-results-rrpe}
%\end{figure*}

%=======================================================================
\section{Conclusions}
%=======================================================================
We have thoroughly investigated multiple algorithms for MLA grid estimation and the influence of the estimation accuracy on light field decoding and calibration of MLA-based light field cameras using the Lytro Illum camera as an example. We proposed a physical camera model and ray tracer implementation to synthesize application-specific white images with known ML centers that we used to quantify the performances of the different algorithms. Based on this camera model, we proposed a new MLA estimation algorithm which explicitly takes into account the natural and mechanical vignetting present in the WI and show that the proposed method outperforms the ones previously discussed in the literature. By releasing the MLA grid estimation pipeline, from WI synthesis to the evaluation of the measured accuracies and making it publicly available, we provide an easy way for the research community to find the optimal estimation parameters for their specific MLA-based application. 

We have shown the importance of accurate MLA grid estimation by decoding the raw sensor images of a Lytro Illum camera as well as its corresponding ray tracer implementation for the different estimation algorithms. With more accurate grid estimates, we find a higher quality in the decoded light fields, mostly in peripheral subaperture images. Additionally, the more accurate grid estimates contribute positively to the camera calibration, lowering the ray reprojection error by up to $\SI{20}{\percent}$, which we have shown for a calibration using line features \cite{Bok2017} as well as a calibration using corner features \cite{Dansereau2013}. Investigations by Bok et al.\ \cite{Bok2017} suggest that increased calibration performance positively effects disparity estimation as well as 3D reconstruction of the calibrated light fields. Furthermore, we assume that measurement application with high accuracy demands, such as light field deflectrometry~\cite{Uhlig2018} or structured light-based methods~\cite{Cai2016, Liu2018a}, can potentially profit from this improved calibration performance. To investigate the impact of calibration performance on a specific measurement task, additional thorough evaluation is necessary.

For the pre-calibration and decoding of light fields, we introduced a new Python package that we make publicly available \cite{git}. The framework offers an easy-to-use user interface as well as a modern modular and object-oriented approach to light field decoding and calibration which is easy to extend for new applications (i.e. new computational cameras). Furthermore, the framework provides modules for general light field and hyperspectral image analysis, such as depth estimation, refocusing, color conversion, and more.

\section{Acknowledgments}
The authors acknowledge support by the state of Baden-Württemberg through bwHPC, a massively parallel computer.
The authors would like to thank David Uhlig for many helpful discussions on the topic.
\vspace{1cm}
%\newpage
{\small
\bibliographystyle{ieee}
\bibliography{lit}
}

\end{document}